\documentclass[twocolumn]{aastex631}
\usepackage{comment}
\shorttitle{Ring Galaxies detected using Convolutional Neural Networks}
\shortauthors{Krishnakumar \& Kalmbach}
\graphicspath{{./}{figures/}}
\usepackage{fancyhdr}
\begin{document}

\title{Analysis of Ring Galaxies Detected Using Deep Learning with Real and Simulated Data}

\author[0000-0001-7950-046X]{Harish Krishnakumar}
\affiliation{Department of Electrical and Computer Engineering\\
Princeton University \\
Princeton, NJ 08544, USA}
\affiliation{Nikola Tesla STEM High School\\
4301 228th Ave NE \\
Redmond, WA 98053, USA}

\author[0000-0002-6825-5283]{J. Bryce Kalmbach}
\affiliation{DIRAC Institute and the Department of Astronomy\\
University of Washington\\
3910 15th Avenue NE, Seattle, WA 98195, USA}



\begin{abstract}
Understanding the formation and evolution of ring galaxies, which possess an atypical ring-like structure, is crucial for advancing knowledge of black holes and galaxy dynamics. However, current catalogs of ring galaxies are limited, as manual analysis takes months to accumulate an appreciable sample of rings. This paper presents a convolutional neural network (CNN) to identify ring galaxies from unclassified samples. A CNN was trained on 100,000 simulated galaxies, transfer learned to a sample of real galaxies, and applied to a previously unclassified dataset to generate a catalog of rings which was then manually verified. Data augmentation with a generative adversarial network (GAN) to simulate images of galaxies was also employed. The resulting catalog contains 1967 ring galaxies. The properties of these galaxies were then estimated from their photometry and compared to the Galaxy Zoo 2 catalog of rings. However, the model's precision is currently limited due to a severe imbalance of rings in real datasets, leading to a significant false-positive rate of 41.1\%, which poses challenges for large-scale application in surveys imaging billions of galaxies. This study demonstrates the potential of optimizing ML pipelines with low training data for rare morphologies and underscores the need for further refinements to enhance precision for extensive surveys like the Vera Rubin Observatory Legacy Survey of Space and Time.

\end{abstract}

\keywords{Ring galaxies (1400), Galaxies (573), Neural networks (1933), Convolutional neural networks (1938), Galaxy evolution (594), Astronomy data analysis (1858), Spectral energy distribution (2129)}


\section{Introduction} \label{sec:intro}

Ring galaxies have been studied for decades due to their unique structures and potential insights into galaxy evolution. Historically, one of the most famous examples, Hoag's Object, was first described by Hoag in 1950, yet its formation remains a mystery despite various hypotheses \citep{1985A&A...153..199B, 1987ApJ...320..454S, 2011}. These galaxies typically feature a luminous core surrounded by a distinct ring of matter. Some ring galaxies, such as the Cartwheel Galaxy, are known to form through galaxy-galaxy collisions \citep{1999IAUS..186...97A, 1996}. Remarkably, the Cartwheel Galaxy hosts an extraordinary number of ultra-luminous X-ray sources, including a potential intermediate-mass black hole \citep{trinchieri2010chandra}. The formation mechanisms of intermediate-mass black holes, which are too massive to result from the collapse of a single star, remain unknown \citep{rose2021formation}. Thus, studying peculiar galaxies like the Cartwheel Galaxy is crucial, and could enhance our comprehension of galaxy formation and evolution.

For these reasons, ring galaxies are of particular interest to study. However, only a limited number of rings have been identified. The Catalog of Southern Ring Galaxies \citep{1995ApJS...96...39B} identified 3692 galaxies south of declination $-17^\circ.$ Most prominent sky surveys, including the Sloan Digital Sky Survey (SDSS; \citealt{2000AJ....120.1579Y}) and the Panoramic Survey Telescope and Rapid Response System (Pan-STARRS; \citealt{chambers2019panstarrs1}), only cover the sky north of around $-30^\circ$, making this data inconvenient for further analysis \citep{Buta2017}.

The Galaxy Zoo 2 project \citep{2013} alleviated this problem through the usage of data from SDSS DR7 \citep{2009ApJS..182..543A}. Morphological classifications for 304,122 galaxies were crowd-sourced from more than 50,000 contributors, and compiled into a database. 3962 of these galaxies were identified to have rings \citep{Buta2017}, though many were poorly resolved, and some were misclassified as a result. Less than $1\%$ of the data, in addition, had tendencies of a collisional ring galaxy \citep{1996}, while most of the data involved a spiral galaxy as a primary component. 

One of the main challenges of relying on human volunteers is the significant amount of time required for data collection and compilation. For instance, the Galaxy Zoo 2 project required over 14 months to gather and process data for 304,122 images. This extended timeframe underscores the limitations of manual classification efforts, especially as the volume of data continues to grow. With millions of images still left to be analyzed and upcoming digital sky surveys such as the Vera Rubin Observatory Legacy Survey of Space and Time (LSST; \citealt{2019ApJ...873..111I}) collecting images of billions of more galaxies, manual classification of morphologies clearly does not suffice for the dramatically increasing volume of data. 

This challenge has inspired previous efforts to classify ring galaxies using computational methods, but these often faced limitations in detection rates and misclassification issues.

The analysis of large samples of galaxies with machine learning, on the contrary, can lead to a greater overall classification accuracy \citep{2011Gan}, with fewer false positives, thus leading to the creation of more comprehensive catalogs of ring galaxies. Particularly, deep learning, and the continuous advancement of convolutional neural networks (CNNs; \citealt{lecun2015deep}) has propelled efficient and accurate image classification.  

\begin{figure*} \label{fig:projection}
\gridline{\fig{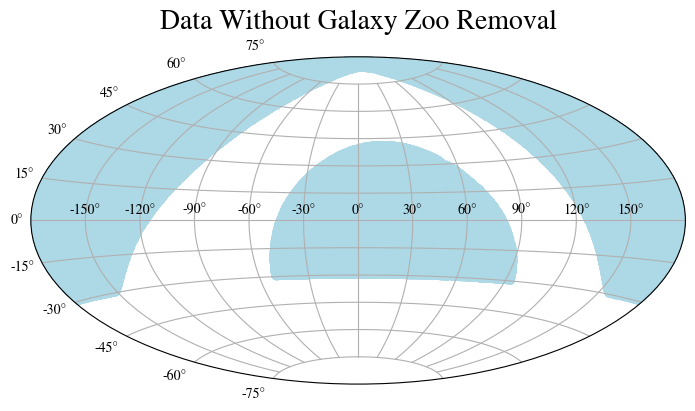}{0.5\textwidth}{(a)}
          \fig{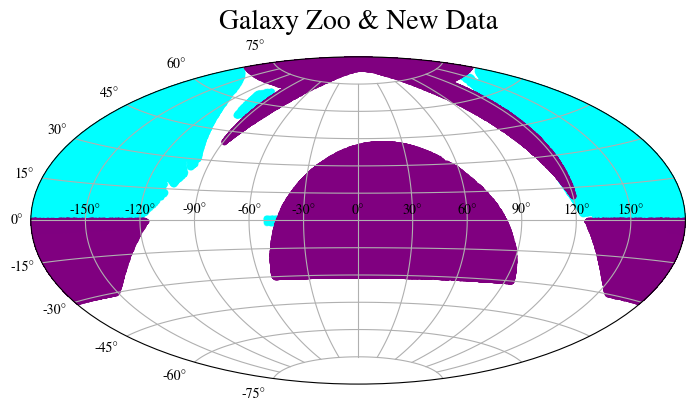}{0.5\textwidth}{(b)}
          }
\caption{Sky coverage of the application data. Left (a): The sky coverage before the subtraction of the sky covered by the Galaxy Zoo 2 project. Right (b): The blue represents the sky coverage of the Galaxy Zoo 2 project while the magenta represents the sky coverage of the modified catalog after removal.}
\end{figure*}

Recently, CNNs have increasingly been used to classify galaxies based on their properties. For instance, \citet{wu2020predicting} used a CNN to predict spectroscopic data from images of galaxies and \citet{2016} used a CNN to distinguish between stars and galaxies. Specifically, the morphology of galaxies can also be accurately classified by these networks. One of the first applications of this was \citet{2015h}, where a CNN was trained to classify the morphology of galaxies in the CANDELS fields with less than a $1\%$ error rate. \citet{2017} described the creation of a CNN to identify gravitational lenses, another unique morphology, trained on a sample of 20,000 simulated galaxies. Building on this concept, \citet{ghosh2020galaxy} trained a CNN to classify galaxies morphologically on a sample of 100,000 simulated galaxies and used a technique known as transfer learning to accurately retrain this model on a small sample of real galaxies, achieving less than a $5\%$ error rate, thus reducing the need for a larger sample size to train an accurate model.

If we want to further understand the formation and evolution of ring galaxies, it is clear that obtaining an expanded catalog of them is essential.CNNs have shown potential in compiling large catalogs of accurately classified rings, provided they are trained and designed appropriately. Particularly, a recent study has detected over 40,000 ring galaxies using a machine learning-based approach, highlighting the potential of computational methods for large-scale galaxy classification \citep{Walmsley_2022}. Classifying galaxies with CNNs is additionally significantly more time efficient than manual analysis \citep{ghosh2020galaxy}, thus showing promise for the analysis of large datasets in hours, compared to the years it would take with manual classification.

In this paper, we present a CNN with the Inception-ResNet-V2 architecture \citep{szegedy2016inceptionv4} to classify ring galaxies \footnote{The code used can be found at \url{https://github.com/harishk30/RingGalaxiesCNNAnalysis/}}. This study aims to create a robust ML pipeline for identifying rare morphologies like ring galaxies, using a limited training dataset of only 3000 galaxies. This model was first trained on a sample of 100,000 simulated ringed and non-ringed galaxies and then transfer learned onto a sample of galaxies classified by Galaxy Zoo 2. Data augmentation was employed via a generative adversarial network (GAN), a deep learning architecture which uses two competing neural networks to simulate images. Then, the model was applied on a sample of $\sim 960,000$ galaxies obtained from the catalog compiled in \citet{2020}. The galaxies identified as rings were manually verified, and compiled into a new catalog. We then examined the specific star formation rate (SSFR), redshift, color and color-mass diagrams of these galaxies, and compared it to a sample of rings from Galaxy Zoo 2 to investigate the distinct properties of the galaxies our model detected. To obtain accurate physical properties of our ring galaxies, we cross-match our catalog with the galSpec \citep{2004MNRAS.351.1151B} database, which provides spectroscopic measurements.

We describe the methods used to obtain the data on which the model is trained, tested and applied on in $\S$ \ref{sec:data}. In $\S$ \ref{sec:model}, we describe the simulation of the initial training data, the selection of a model architecture, the training of the model, and the application of transfer learning. In $\S$ \ref{sec:prop}, we describe obtaining properties of the newly obtained catalog. In $\S$ \ref{sec:results}, we present the expanded catalog of ring galaxies and their properties. Finally, in $\S$ \ref{sec:discussion} and $\S$ \ref{sec:conclusion} we discuss the results and their potential applications.

\section{Obtaining Data} \label{sec:data}

This paper primarily discusses the usage of a CNN to detect ring galaxies in real data sets. A catalog of ring galaxies was first obtained from the Galaxy Zoo 2 project (\citealt{2013}) to train the model. 3962 galaxies were classified as "clean" samples of rings. As Galaxy Zoo 2 participants were shown images from the Dark Energy Camera Legacy Survey (DECaLS), images to train the model were obtained from the DESI Legacy Imaging Surveys \citep{2019d} as it includes DECaLS as a primary component.

Cutouts of the training images were 256 pixels on a side. As the galaxies were arranged in descending order by angular diameter, the resolution of the images was iterated from 0.65" to 0.09" per pixel to ensure that the galaxies occupied equal areas in the frame.

\begin{figure}
\plotone{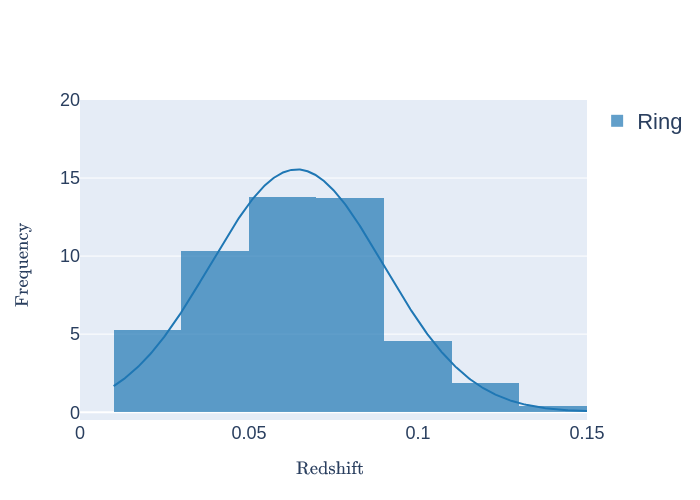}
\caption{Redshift distribution for the Galaxy Zoo 2 rings.}
\label{fig:gzr}
\end{figure}

\begin{figure}
\plotone{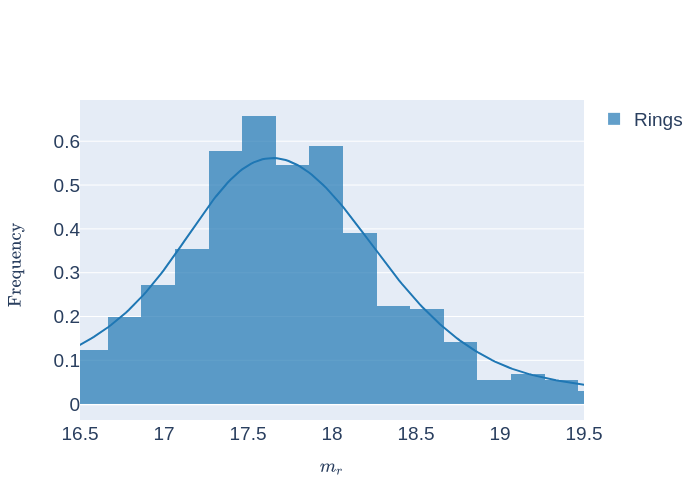}
\caption{R-band magnitude distribution for the Galaxy Zoo 2 rings.}
\label{fig:mr}
\end{figure}

The training sample of galaxies was then manually verified. Only about 3117, or 78.6\%, of the initial sample of ring galaxies, was found to visually exhibit characteristics of a ring. The morphologies of contaminant galaxies were primarily elliptical galaxies and barred spiral galaxies. As described in $\S$ \ref{sec:model}, 80\% of this data was reserved for training while 20\% was reserved for testing. The redshift distribution of this sample is shown in Figure \ref{fig:gzr}, and the r-band magnitude distribution of the sample is shown in Figure \ref{fig:mr}.

The model was applied to detect rings in a catalog of galaxies identified in the Pan-STARRS survey \citep{2020g}. This catalog will henceforth be referred to as the Goddard-Shamir catalog. The Goddard-Shamir catalog initially contained $1,685,349$ galaxies. Many of these galaxies had not been previously classified, offering an opportunity to demonstrate the model's effectiveness on a streamlined catalog. Due to the overlapping sky coverage with Galaxy Zoo 2, which served as the training set for the model, we excluded the area covered by Galaxy Zoo 2 from the Goddard-Shamir catalog. This adjustment yielded a final catalog of $966,329$ galaxies, which allowed for an evaluation of the model on a substantial dataset of previously unclassified galaxies. Images for these galaxies were sourced from the DESI Legacy Imaging Surveys, following the same procedures as for the training data.

\section{Creating \& Training the Convolutional Neural Network} \label{sec:model}

\begin{figure} 
\begin{interactive}{js}{interactive.tar.gz}
\plotone{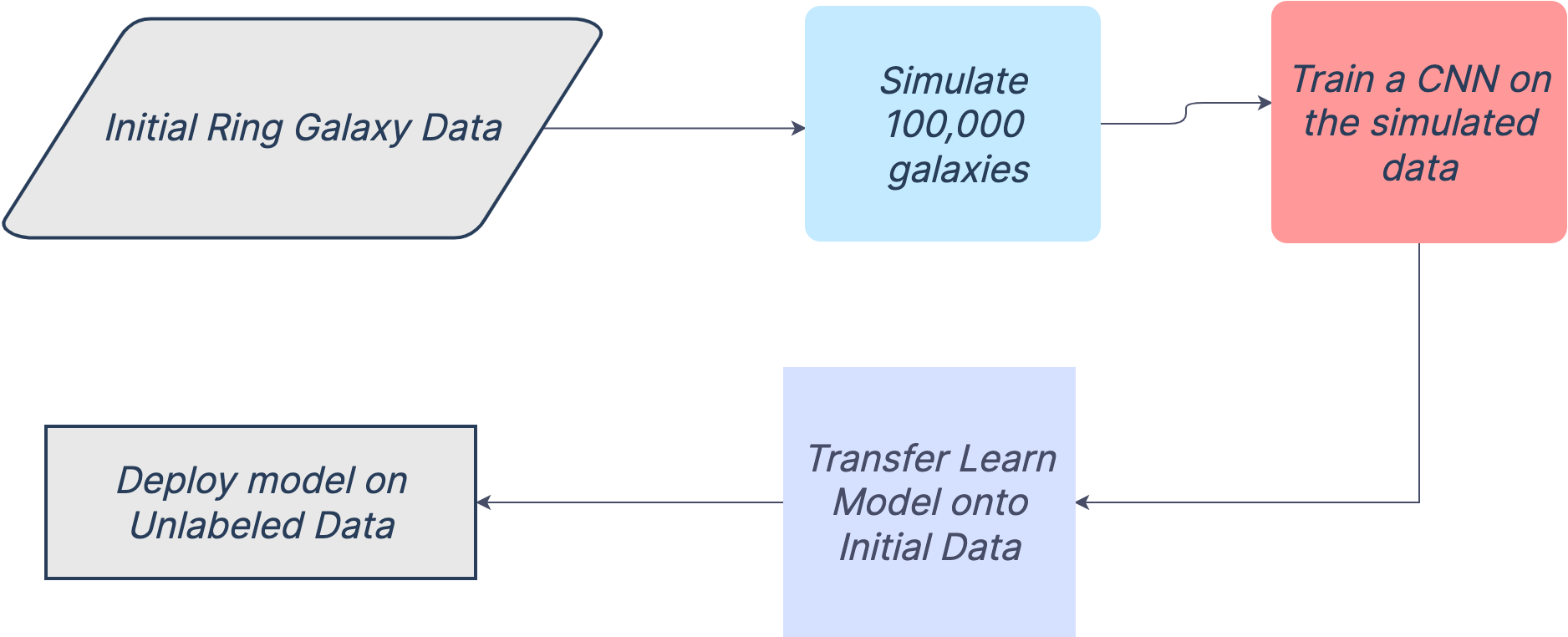}
\end{interactive}
\caption{Workflow starting from the initial data to the final CNN. Obtaining the initial data is described in $\S$ \ref{sec:data}. The simulation of 100,000 galaxies is described in $\S$ \ref{subsec:sim}. The CNN trained on the simulated data is described in $\S$ \ref{subsec:CNN}. Finally, the process of transfer learning with the CNN is described in $\S$ \ref{subsec:tl}. }
\label{fig:interactive}
\end{figure}

To train an accurate CNN, large data sets are typically required. However, with existing catalogs of ring galaxies being limited to only a few thousand galaxies, a different approach is required to obtain the most accurate model. A technique known as transfer learning \citep{zhuang2020comprehensive} is used to first train the model on a sample of simulated galaxies, and the information learnt is used to retrain the model on the smaller sample of real galaxies. The steps used to train the model are shown in in Figure \ref{fig:interactive}.

\begin{figure*} \label{fig:simulation}
\gridline{\fig{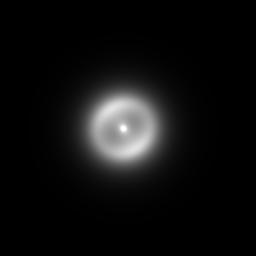}{0.3\textwidth}{(a)}
          \fig{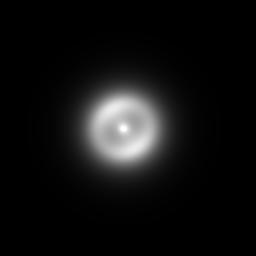}{0.3\textwidth}{(b)}
          \fig{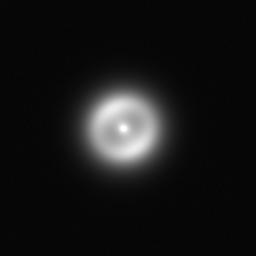}{0.3\textwidth}{(c)}
          }
\caption{A simulated ring galaxy and modifications to make it further resemble real data. Left (a): The simulated light profile initially created by {\tt\string GALFIT}. Middle (b): The simulated light profile after convolution with a point spread function (PSF). Right (c): The simulated light profile after convolution with a PSF and the addition of randomized noise.}
\end{figure*}
\subsection{Simulating Galaxies\label{subsec:sim}}
To simulate the initial training sample of ring galaxies, the {\tt\string GALFIT} \citep{2002} program was used on an Ubuntu virtual machine. {\tt\string GALFIT} is typically used to fit light profiles to astronomical objects, but in this context, as in \citealt{ghosh2020galaxy}, it is used to create light profiles for randomized galaxies to be used as training data. 

The training data set contained 200,000 simulated galaxies, with an even split between ringed and non-ringed galaxies. The galaxies were simulated using a Sérsic profile \citep{2010p}, which describes how the intensity of a galaxy varies with distance from its center. The functional form of the Sérsic profile is described by
\begin{equation} \label{eq:sersic}
\Sigma(r) = \Sigma_e \exp \left[-\kappa\left(\left(\frac{r}{r_e}\right)^{1/n}-1\right)\right].
\end{equation}$\Sigma_e$ represents the pixel surface brightness at the radius $r_e$, which represents the radius at which half of the galaxy's flux is contained. The Sérsic index of the galaxy, $n$, controls where the light of the galaxy is concentrated, and $\kappa$ is dependent on this parameter to ensure that half of the flux stays within radius $r_e$ from the center. 

However, a ringed structure cannot accurately be represented by one standard Sérsic function. As ring galaxies have a pronounced gap in their center, the initial  Sérsic function needs to be multiplied by a hyperbolic tangent truncation function. As described in \citealt{2010p}, this is given by:

\begin{equation} \label{eq:trunc}
0.5\left ( \tanh\left [ \left ( 2 - B \right )\frac{r}{r_{break}} + B\right ] + 1\right )
\end{equation}

where\begin{equation} \label{eq:B}
B = 2.65-4.98\left ( \frac{r_{break}}{r_{break} - r_{soft}} \right ).
\end{equation}

In the function, $r_{break}$ represents the radius at which $99\%$ of the galaxy's flux is enclosed while $r_{soft}$ represents the radius at which $1\%$ of the galaxy's flux is enclosed. However, the inner core of the galaxy still needed to be represented, thus requiring a second Sérsic profile without a truncation function applied. When simulating the sample of non-ringed galaxies, only a single Sérsic profile was needed.

\begin{deluxetable*}{cchccccc}
\tablenum{1}
\tablecaption{Parameters to Simulate Ringed and Non-Ringed Galaxies}
\label{table:1}
\tablewidth{0pt}
\tablehead{
\colhead{Component Name} & \colhead{Sérsic Index} & \nocolhead{Common} & \colhead{Half-Light Radius} &
\colhead{Axis Ratio} 
& \colhead{Integrated Magnitude}
& \colhead{Break Radius} 
&  \colhead {Softening Radius} \\
\colhead{} & \colhead{} & \nocolhead{Name} & \colhead{(Pixels)} &
\colhead{} & \colhead{(AB)} & \colhead{(Pixels)} &
\colhead{(Pixels)}
}
\decimalcolnumbers
\startdata
\multicolumn{8}{c}{\emph{Ring Galaxy}} \\
Outer Ring  & 1.5 - 3.0 & & 2.0 - 8.0 & 0.7 - 1.0 & 15.0 - 20.0 & N/A & N/A\\
Truncation Function & N/A & N/A & N/A & 0.7 - 1.0 & N/A & 40.0 - 70.0 & 20.0 - 40.0\\
Inner Core & 0.5 - 4.0 & & 5.0 - 15.0 & 0.5 - 1.0 & 14.0 - 18.0 & N/A &  N/A\\
\multicolumn{8}{c}{\emph{Non-Ringed Galaxy}} \\
Sérsic Profile & 0.0 - 8.0 && 55.0 - 88.0 & 0.1 - 1.0 & 17.0 - 25.0 & N/A & N/A \\
\enddata
\end{deluxetable*}

To determine the general parameters to simulate ring galaxies, light profiles are fit to ten randomly selected galaxies from the sorted Galaxy Zoo 2 catalog using {\tt\string GALFIT}. The parameters derived from these fits are used to create a distribution of values for simulating ring galaxies. This approach ensures a representative sample of ring galaxy parameters.

The parameters for simulating ring galaxies are then randomly generated based on the distributions obtained from the ten galaxy fits. Parameters for non-ringed galaxies are also randomly generated with a uniform distribution. These parameters are described in Table \ref{table:1}. The position angle of the galaxies is iterated between $-90^\circ$ and $90^\circ$.

The {\tt\string GalaxySim}\footnote{\url{https://github.com/aritraghsh09/GalaxySim}.} code \citep{ghosh2020galaxy} was used with modifications customized to the ringed galaxies parameters to create the input files for {\tt\string GALFIT}, run {\tt\string GALFIT} on the input files, and modify the generated images to be more realistic.

After creating the input files, {\tt\string GALFIT} is run to create simulated light profiles. To make the simulations more realistic, we incorporate background noise and PSF effects from real observations. Specifically, we use data from the DESI Legacy Imaging Surveys Data Release 8 (DESI DR8).

The images are convolved with a point spread function (PSF) obtained from DESI DR8, representing the dispersion of light caused by atmospheric disturbances and optical limitations. For the background noise, we use {\tt\string SourceExtractor} \citep{1996A&AS..117..393B} to mask galaxies in DESI DR8 images. The remaining background noise from these masked images is then injected into our simulated galaxies. This approach ensures that our simulated images closely mimic the noise characteristics of real observations.

An example of these modifications is shown in Figure \ref{fig:simulation}.

Following this, a Python script is created to first scale the images, keeping $99.5\%$ of the initial pixels, and then convert them from \textit{FITS} to \textit{JPG} files to train the model on, as this is the file format the cutouts of the real galaxies are in.

This approach to simulation, using multiple prototype galaxies and incorporating real observational effects, provides a robust and realistic training dataset for our model.

\subsection{Training the Network\label{subsec:CNN}}

To train an accurate CNN, large data sets are typically required. However, with existing catalogs of ring galaxies being limited to only a few thousand galaxies, a different approach is required to obtain the most accurate model. A technique known as transfer learning \citep{zhuang2020comprehensive} is used to first train the model on a sample of simulated galaxies, and the information learnt is used to retrain the model on the smaller sample of real galaxies. The steps used to train the model are shown in Figure \ref{fig:interactive}.

\begin{deluxetable*}{cchccccc}
\tablecaption{Metrics for Various Models with Simulated Data}
\tablenum{2}
\label{table:2}
\tablewidth{0pt}
\tablehead{
\colhead{Model Name} & \colhead{Accuracy} & \nocolhead{FPR} & \colhead{Precision} &
\colhead{Recall} 
& \colhead{$F_{1}$ Score}
& \colhead{MCC} 
&  \colhead {AUC} 
}
\decimalcolnumbers
\startdata
ResNet-101 & $97.28\%$ && 1.000 & 0.939 &  0.968 & 0.946 & 0.993\\
Inception-ResNet-V2 & $98.62\%$ & & 0.981 & 0.988 & 0.984 & 0.972 & 0.997\\
\enddata
\tablenotetext{a}{The given metrics represent the following information: accuracy is the fraction of predictions which the model correctly identified, precision is the fraction of true positives identified relative to all positive identifications, recall is the fraction of positive samples identified correctly, $F_{1}$ score is the balance between precision and recall, MCC measures the overarching quality of the binary classifier, and AUC represents the balance between true positive rate and false positive rate.}
\end{deluxetable*}

\begin{deluxetable*}{cchccccc}
\tablecaption{Metrics for Various Transfer Learned Models}
\tablenum{3}
\label{table:3}
\tablewidth{0pt}
\tablehead{
\colhead{Layers Unfrozen} & \colhead{Accuracy} & \nocolhead{filler} & \colhead{Precision} &
\colhead{Recall} 
& \colhead{$F_{1}$ Score}
& \colhead{MCC} 
&  \colhead {AUC} 
}
\decimalcolnumbers
\startdata
Output & $ 78.5\%$ & &  0.766 &  0.766 &  0.766 &  0.567 &  0.897 \\
Block B, C, Output &  $93.9\%$ &&  0.904 &  0.971 &   0.936 &  0.881 &  0.972 \\
All (No Transfer Learning) &  $85.0\%$ & &  0.823 & 0.702 &  0.758 &  0.655 & 0.884\\
\enddata
\end{deluxetable*}

Using the {\tt\string Keras} API, which is built on {\tt\string TensorFlow}, the model was created. Two models were initially tested and evaluated. The first had the ResNet-101 architecture, which has 101 layers and utilizes residual connections. The second had the Inception-ResNet-V2 architecture \citep{szegedy2016inceptionv4}, which has 164 layers and utilizes a combination of both inception modules and residual connections. The inception modules are organized in three blocks, A, B, and C, and each has a series of repeating layers. These networks were trained on the $100,000$ simulated images for 11 epochs with a batch size of 32 and a learning rate of $0.0001$.

To evaluate these networks, $80\%$ of this data was reserved for training, while $20\%$ was reserved for testing. Half of the data reserved for testing was used for validation, while the other half was used as a true test set, which avoids any influence the process of training might have had. The results of the evaluation are described in Table \ref{table:2}. Using a holistic evaluation of the various metrics, Inception-ResNet-V2 was clearly the best network for the classification of the simulated ring galaxies. As a result, this network was then used for the process of transfer learning onto the real data.

For more details about CNNs and our specific architecture, please refer to Appendix \ref{sec:anncnn}.

\subsection{Transfer Learning \& GANs\label{subsec:tl}}

To transfer the information the model learnt from the simulated data to the real data, a technique known as transfer learning \citep{zhuang2020comprehensive} was used. To get better accuracy when limited data is available, transfer learning helps as the model is pre-trained on a different data set before training on the desired data. 

Transfer learning involves freezing particular layers of the model, typically all layers excluding the output layers, and retraining the layers which remain unfrozen. The earlier layers most likely do not need retraining, as they are primarily meant to extract trivial features. A common example of transfer learning is when a model trained on the ImageNet data set is applied to a new data set. However, similar to \citealt{ghosh2020galaxy}, this model is first trained on a sample of simulated galaxies resembling the real data, and then transfer learned onto the real data. 

To get the best possible model, different variations of transfer learning were tested. This included retraining just the output layers, retraining inception blocks B, C, and the output layers, and retraining the entire model without using transfer learning. 

\begin{figure*} 
\plotone{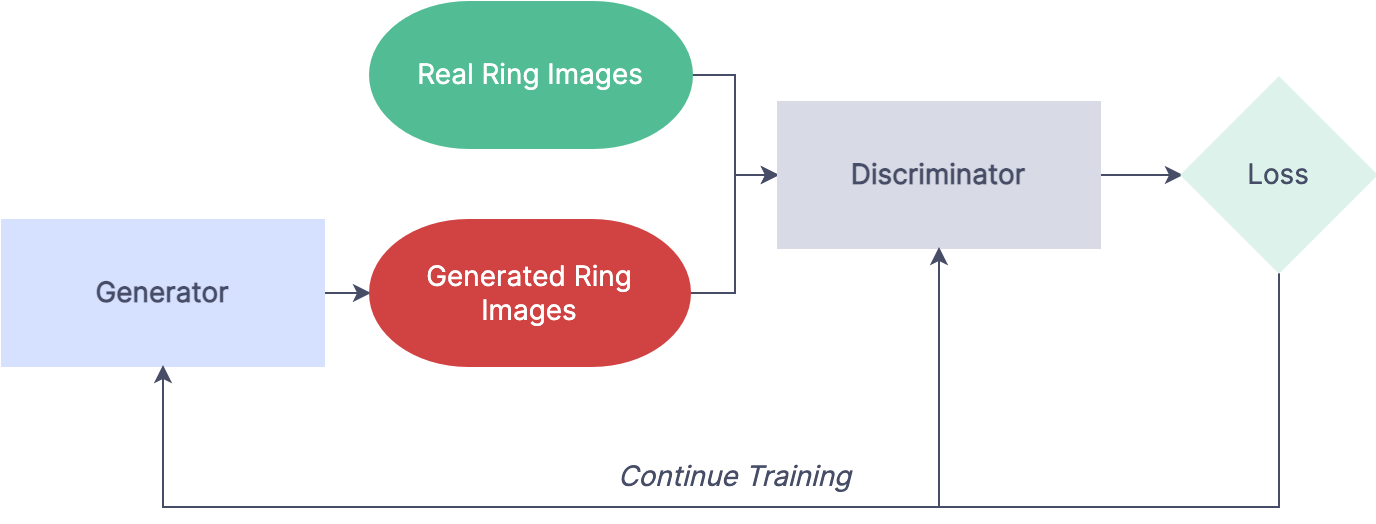}
\caption{An illustration of the architecture of a basic GAN with a discriminator and a generator.}
\label{fig:gan}
\end{figure*}

The training data set contained about 9318 images, 3117 of which were ring galaxies. This data set was intentionally designed with a 2:1 ratio of non-ringed galaxies to ringed galaxies to help alleviate the data imbalance and ensure minimal contamination in the final sample of ringed galaxies obtained when applied to an unclassified data set. Data augmentation was used to prevent overfitting, along with early stopping, which stopped training once convergence was reached. Particularly, a generative adversarial network (GAN), the general architecture of which is shown in Figure \ref{fig:gan}, was used for data augmentation. GANs are generative models which utilize two competing neural networks: a generator is trained to generate fake data, while a discriminator is employed to distinguish generated images from real samples. As a GAN trains, the generator improves at generating fake data, and the accuracy of the discriminator starts to decrease. Due to the ability of GANs to generate realistic images, they can be used to enlarge an existing dataset.

\begin{figure}
\plotone{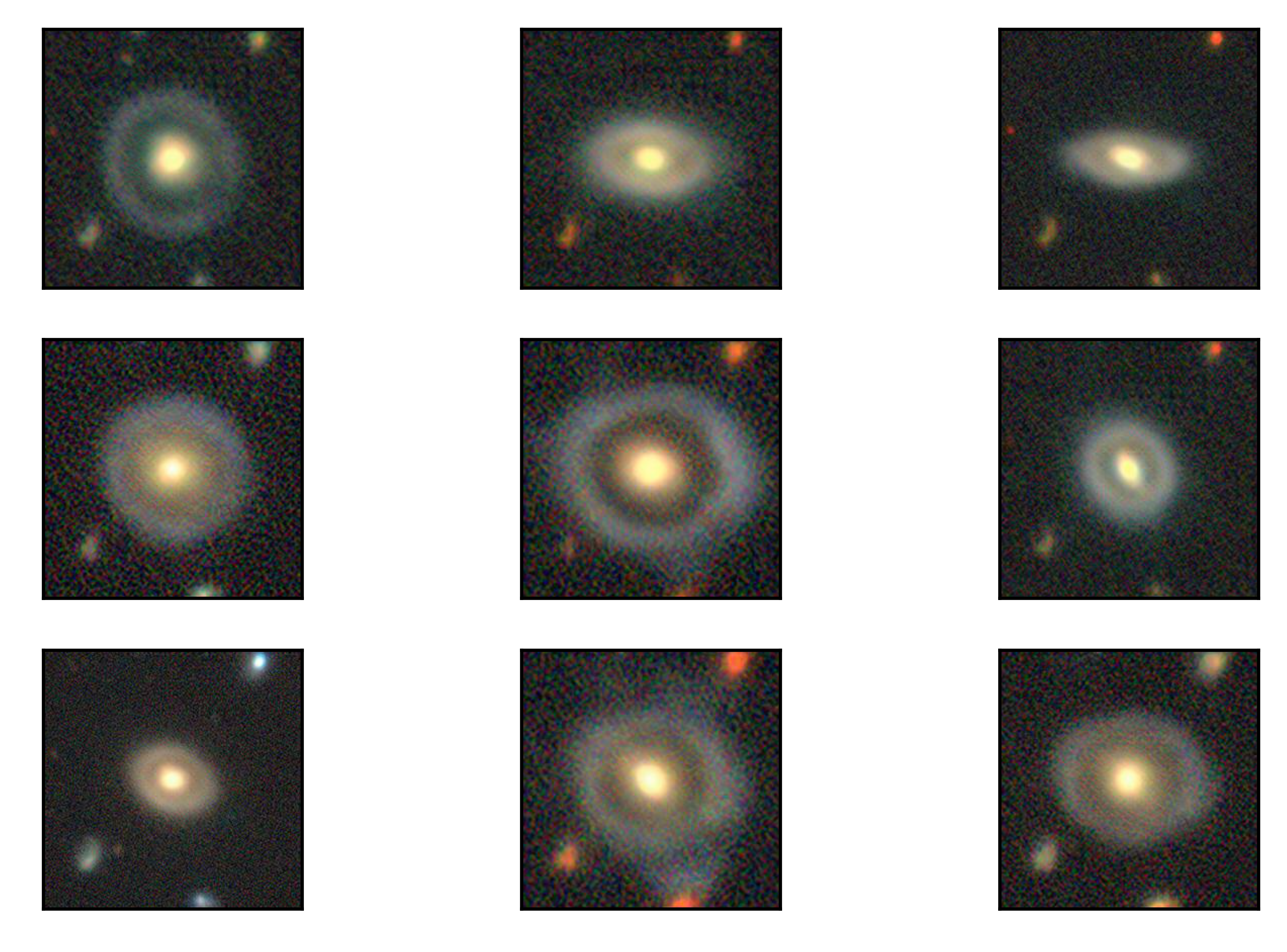}
\caption{Nine of the GAN's 20,000 generations. The identical background sources are discussed in $\S$ \ref{sec:discussion}.}
\label{fig:gaen}
\end{figure}

Three different GAN architectures were tested out. This included StyleGAN-2-Ada \citep{karras2020training}, ProGAN \citep{karras2017progressive} and DcGAN \citep{radford2015unsupervised}. DcGAN, the simplest out of the three, uses deep convolutional neural networks as both the discriminator and the generator. ProGAN starts from small, 4x4, images, and the model eventually "grows" until it can generate 256x256 images. StyleGAN-2 is an extension of ProGAN, allowing more detailed features to be recreated in its generations. ADA, or adaptive discriminator augmentation, allows StyleGAN-2 to train without overfitting on small datasets. Out of the three, StyleGAN-2-Ada consistently generated the highest quality images, of which nine images out of the 20,000 generations used are shown in Figure \ref{fig:gaen}. 

Additionally, a differential learning rate was used. The output layers, which were completely being retrained, were assigned a learning rate of $10^{-3}$. However, the layers behind the output layers, if retrained, were assigned a learning rate of $10^{-5}$. After the model reached convergence, an extra fine-tuning step was performed. This involved unfreezing the entire model, and retraining it at the extremely low learning rate of $10^{-7}$, such that it could make minor adjustments.

The model was trained with a batch size of 32 and the early stopping had a patience of 10 epochs. The model was again trained on $80\%$ of the initial data, with $10\%$ being reserved as the validation data set and $10\%$ being reserved as the test data set. The results of the transfer learning are described in Table \ref{table:3}. The model where both inception block B, C and the output layers were retrained was chosen because it performed best on every metric. Then, this model was run on the unlabeled data. The galaxies classified as rings are described in $\S$ \ref{sec:results}. 

\section{Inferring the properties of the galaxies} \label{sec:prop}
To determine the properties of the identified ring galaxies, we cross-match our catalog with the galSpec \citep{2004MNRAS.351.1151B} database. We only consider galaxies with available spectra in galSpec for our analysis. From these matched galaxies, we obtain key parameters directly from galSpec: stellar mass ($M_*$), specific star formation rate (SSFR), and redshift.

The cross-matching process involves comparing the coordinates of our ring galaxies with those in the galSpec database. We use a matching radius of 2 arcseconds to account for potential small differences in the reported positions.

By focusing solely on the spectroscopically-confirmed subset of our ring galaxy catalog, we ensure a high-quality dataset for our analysis. This approach, while potentially reducing our sample size, provides the most reliable measurements of physical properties for these unique objects.

In $\S$ \ref{sec:results} and $\S$ \ref{sec:discussion}, we will describe these results and how they can be used to infer the formation and evolution of these galaxies.

\begin{figure*}
\includegraphics[width=\textwidth]{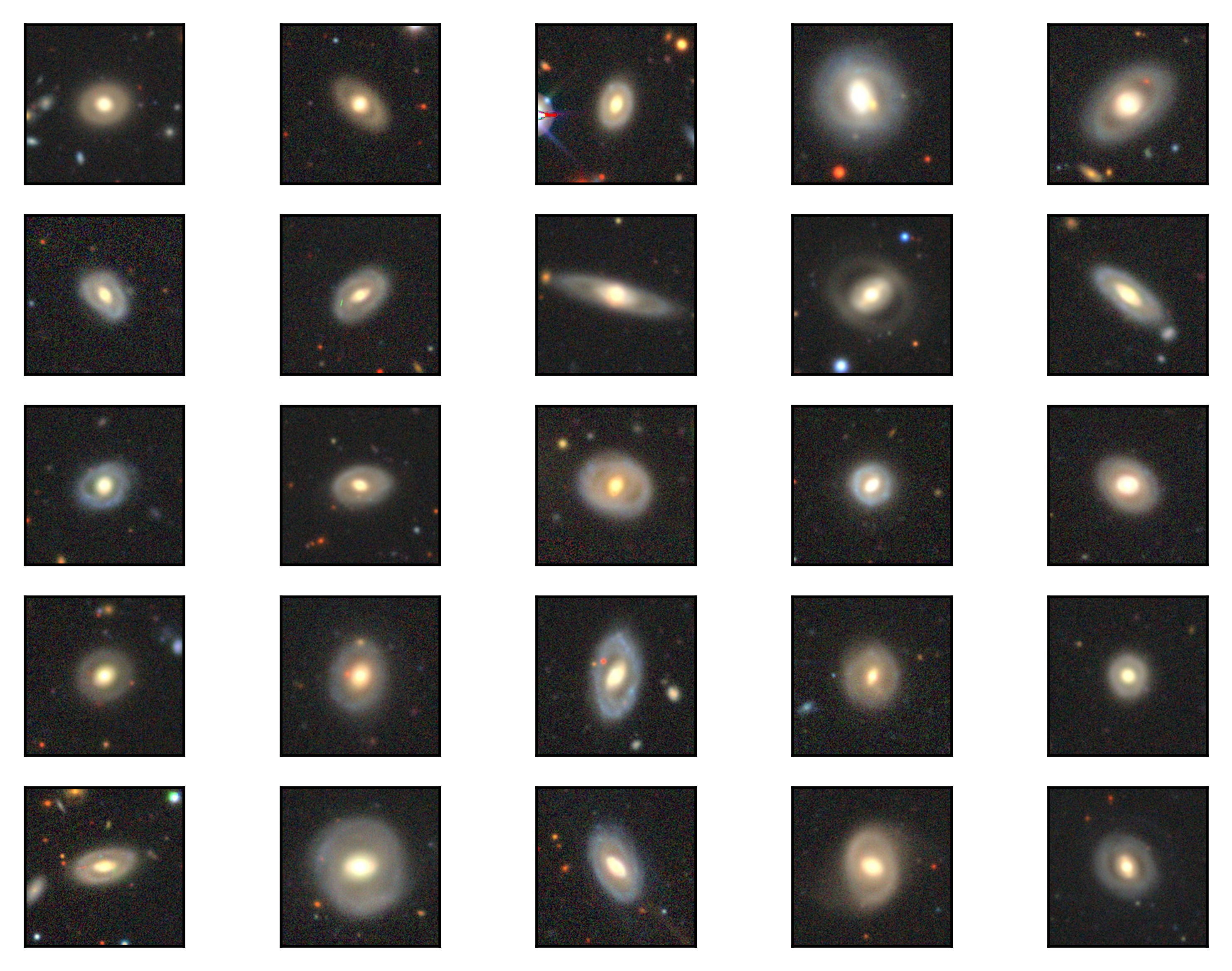}
\caption{25 galaxies from the catalog of 1967 rings.}
\label{fig:rings}
\end{figure*}

\section{Results} \label{sec:results}

\subsection{Classification of Unclassified Data\label{subsec:cl}}

The model was applied to 966,329 previously unclassified galaxies from the Goddard-Shamir catalog after training. Images for these galaxies were obtained from the DESI Legacy Imaging Surveys. A cutoff of a $99.5\%$ probability to classify a galaxy as a ring was applied on the output from the final sigmoid layer of the model to ensure that the least number of non-ringed galaxies were misclassified as rings.

After the classifications of the model were manually filtered, 1967 ring galaxy candidates were found. These galaxies have not previously been classified as rings, making this the first time they have been recognized as rings. Figure \ref{fig:rings} shows images of 25 galaxies from this catalog. Additionally, the application of this model to the $\sim$950,000 previously unclassified galaxies took only around 10 hours, markedly better than the 14 months it took Galaxy Zoo 2 volunteers to sort through the $\sim $300,000 galaxies of SDSS DR7.

The unclassified data set differed from the training, validation and confirmation data in that it was not specifically filtered to include a certain proportion of non-ringed and ringed galaxies: in a "real" data set, less than $0.1\%$ of the galaxies are rings, while the model was trained with a 2:1 ratio of non-ringed galaxies to rings. This led to a larger proportion of non-ringed galaxies being misclassified relative to ringed galaxies: 1357 non-ringed galaxies were misclassified. 

\begin{figure} 
\centering
\includegraphics[width=40mm,scale=0.3]{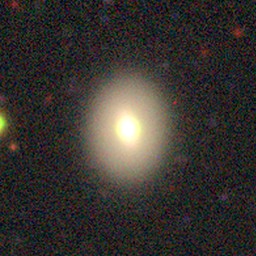}
\caption{A galaxy misclassified as a ring.}
\label{fig:misclass}
\end{figure}

Many of the model's misclassifications tended to be small elliptical galaxies, like the one shown in Figure \ref{fig:misclass}. The model likely identified a common feature in these galaxies which it misidentified as ringed structure, causing it to misclassify them.

\subsection{Analysis of Properties\label{subsec:prop}}

The stellar mass ($M_*$), specific star formation rate (SSFR), and redshift of the newly identified rings were determined by crossmatching galaxies which had available spectra from galSpec \citep{2004MNRAS.351.1151B}. Additionally, the apparent magnitudes of the galaxies were determined by
\begin{equation}
m = -2.5 \cdot \log_{10}{\left(\frac{F}{F_0}\right)},
\end{equation}
where $F$ is the flux density of the galaxy through a given filter and $F_0$ is the zero point flux density through that filter. Different colors of the galaxies, including FUV - r and g - r color, could be obtained by subtracting the apparent magnitudes of the galaxies in those filters in their respective combinations.

We compared our sample to a sample of 1054 rings from Galaxy Zoo 2. These were chosen to see how our catalog of rings differed from previous catalogs.  

Using the Kolmogorov-Smirnov (KS) test, the distribution of SSFRs in our sample of ring galaxies was compared to that of the control sample of previously identified rings. The null hypothesis that both were from the same distribution was rejected with a p-value of .002, less than the p-value threshold of .05 which we had chosen. 

\begin{figure} 
\centering
\includegraphics[width=90mm,scale=0.3]{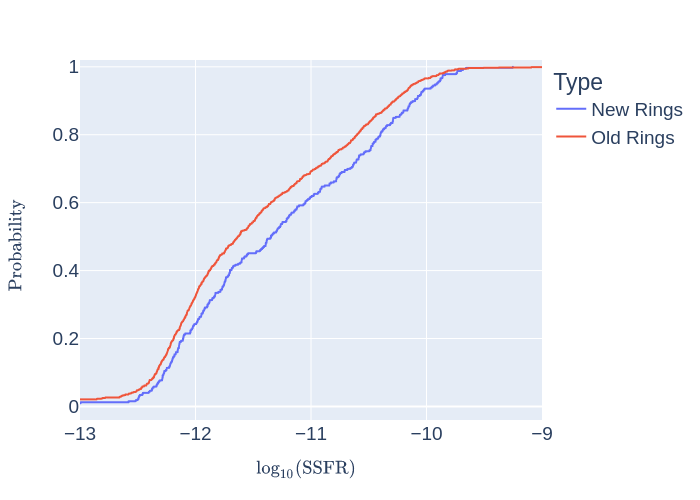}
\caption{The eCDF of the newly identified sample of rings and the Galaxy Zoo 2 rings.}
\label{fig:eCDF}
\end{figure}

The KS-test compares the empirical distribution function (eCDF) of both samples of galaxies. Looking at the eCDFs of the galaxies in Figure \ref{fig:eCDF}, it is clear that the eCDF of the Galaxy Zoo 2 rings is shifted left, indicating that the probability of finding rings from Galaxy Zoo 2 at lower SSFR values is higher than from our catalog. 

After this, the color-mass diagrams of both samples were compared against each other. On the color-mass diagram, redder galaxies (which have a lower apparent magnitude in the r-band filter) are typically older and less star forming than those which are bluer.

\begin{figure} 
\centering
\includegraphics[width=90mm,scale=0.3]{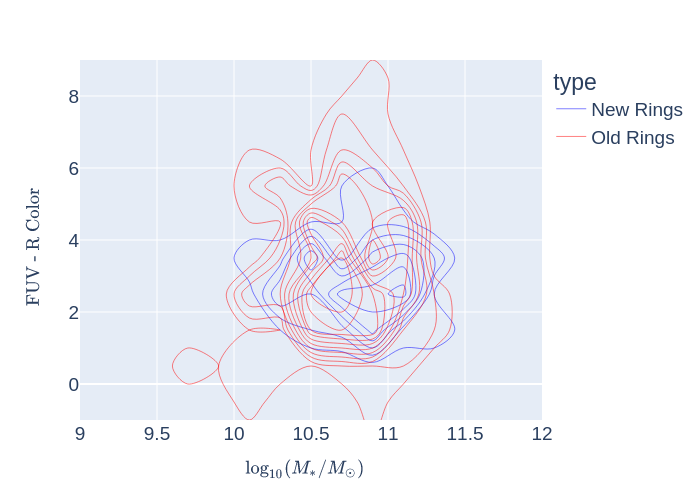}
\caption{Color-mass diagram for FUV - r color for the newly identified sample of rings and the Galaxy Zoo 2 rings.}
\label{fig:FUVr}
\end{figure}

The FUV - r color of both our sample and the Galaxy Zoo 2 sample of rings were plotted against $\log_{10}(M_{*} / {M}_\odot)$ on a density contour plot in Figure \ref{fig:FUVr}. The Galaxy Zoo 2 rings were seen to extend into higher, or redder, values, at intermediate masses. However, this effect was not seen as prominently at other wavelengths. While other colors, such as g - r and NUV - r color, were found, the most pronounced difference was seen in FUV - r color. 

A frequency histogram of the masses of both samples of rings is shown in Figure \ref{fig:mass}. The newly identified sample of rings is shown to peak at a slightly higher value of mass than the Galaxy Zoo 2 rings.

Then, a frequency histogram of the magnitudes in the r-band filter of both samples was plotted and is shown in Figure \ref{fig:magnitude}. The Galaxy Zoo 2 rings extend into higher magnitudes when compared with the newly identified sample.

\section{Discussion} \label{sec:discussion}
\subsection{Model \& Applications}
When applied to the unclassified data set, the model identified 1967 ring galaxy candidates after manual verification. The model achieved a $58.9\%$ precision and a false-positive rate of $41.1\%$ when applied to unclassified data and detected 2035.5 ring galaxies per 1 million. This precision demonstrates the potential of convolutional neural networks (CNNs) in identifying rare morphologies even with extremely limited training data.

\begin{figure} 
\centering
\includegraphics[width=90mm,scale=0.3]{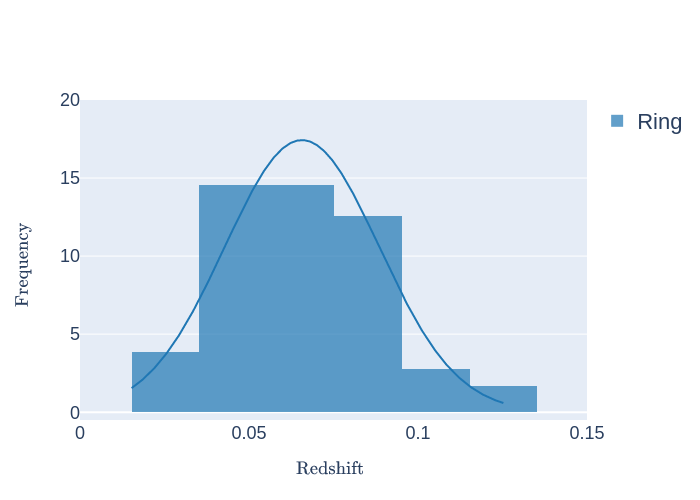}
\caption{Frequency histogram of redshifts for newly identified rings.}
\label{fig:rs}
\end{figure}

Finally, the frequency of the redshifts of the newly identified rings were plotted as a histogram, as seen in Figure \ref{fig:rs}. 

Ring galaxies are particularly challenging to detect due to their diverse morphologies: some are face-on while others lie nearly edge-on. Additionally, many features that resemble rings may not indicate actual ring galaxies. This complexity was reflected in the misclassified sample of galaxies, which mostly included small elliptical galaxies (such as in Figure \ref{fig:misclass}), mergers of two galaxies, and occasionally small spiral galaxies. Although the model avoided misclassifying images with noise and artifacts, it struggled with certain galaxies exhibiting ring-like traits.

\begin{figure} 
\centering
\includegraphics[width=90mm,scale=0.3]{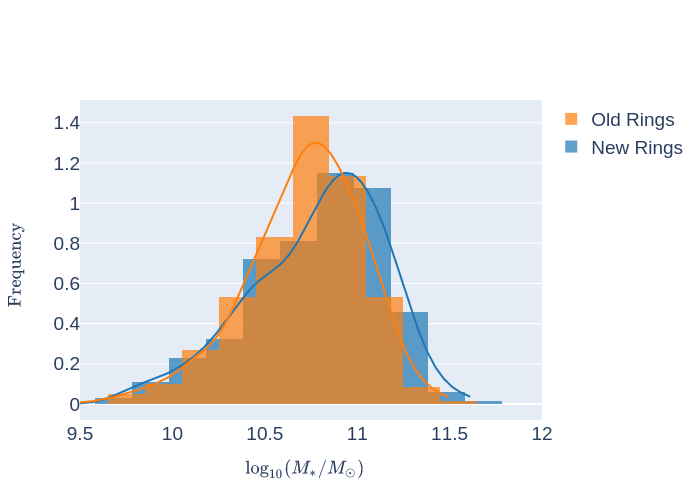}
\caption{Frequency histogram of masses for the newly identified sample of rings and the Galaxy Zoo 2 rings.}
\label{fig:mass}
\end{figure}

\begin{figure} 
\centering
\includegraphics[width=90mm,scale=0.3]{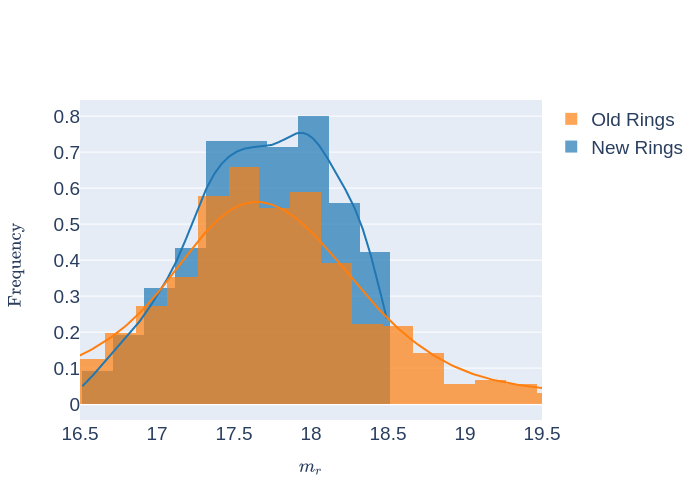}
\caption{Frequency histogram of magnitudes in both samples of rings.}
\label{fig:magnitude}
\end{figure}

One potential improvement to enhance the accuracy of the model could involve increasing the variety of the initial simulated sample of galaxies to include collisional and Hoag-type ring galaxies, in addition to standard rings. Multiple CNNs could also be stacked in an "ensemble" network to ensure the best-supported prediction for each image.

Furthermore, leveraging outlier detection methods could significantly improve the model's robustness, particularly in the context of extreme data imbalance where less than 0.1\% of galaxies are rings. Outlier detection techniques identify unusual data points that deviate from the norm, helping to isolate the rare ring-like structures amidst a vast majority of non-ringed galaxies. By incorporating these techniques, the model could be fine-tuned to better distinguish between true ring galaxies and false positives.

Active learning algorithms, such as Astronomaly \citep{Lochner_2021}, present another promising avenue for enhancing the model's performance in dealing with data imbalance. Astronomaly uses a human-in-the-loop approach to iteratively refine the model's accuracy. By incorporating expert feedback on the model's predictions, the system can dynamically improve its classification capabilities. This approach is particularly advantageous for identifying rare morphologies like ring galaxies, where expert validation is crucial. The active learning framework helps prioritize the most informative samples for labeling, thus addressing the imbalance by focusing on the minority class more effectively.

\begin{figure} 
\centering
\includegraphics[width=90mm,scale=0.3]{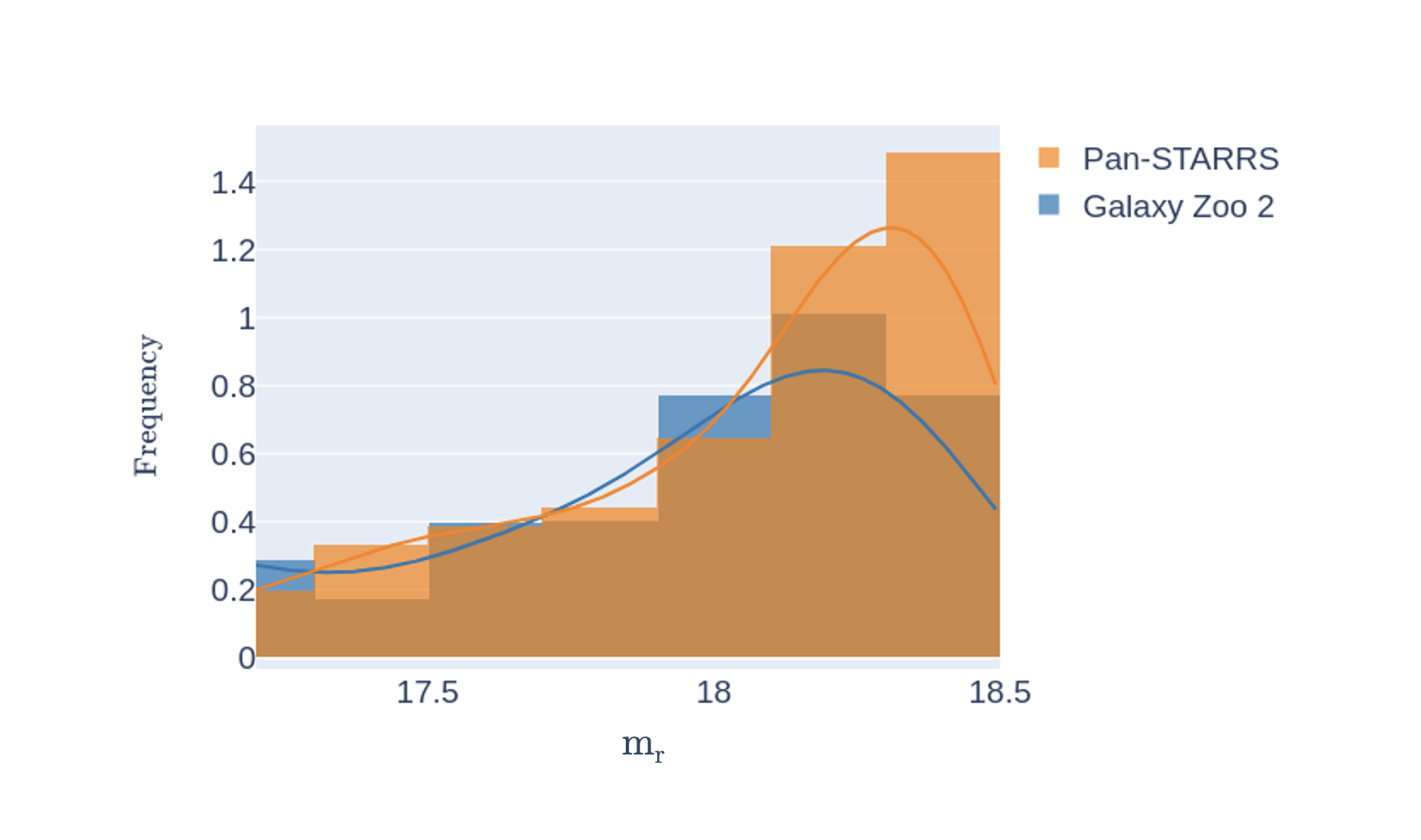}
\caption{Frequency histogram of magnitudes for a random sample of around 1000 galaxies from the Galaxy Zoo 2 catalog and the Goddard-Shamir catalog.}
\label{fig:magc}
\end{figure}

The generative adversarial network (GAN) used for data augmentation also presents opportunities for improvement. Many of the GAN's generated images, as seen in Figure \ref{fig:gaen}, have similar background sources, potentially affecting the model's accuracy. Future iterations could improve accuracy by masking out background sources in training images, ensuring that the GAN focuses on simulating realistic ring structures without extraneous noise.

These results underscore that a CNN can be developed to detect other peculiar morphologies, such as green pea galaxies, even with limited training data. This research demonstrates that methods like transfer learning, paired with synthetic data and GANs, can achieve competitive accuracy.

However, to apply similar models to large-scale surveys like the LSST, it is crucial to improve the model's precision. The LSST is expected to catalog approximately 10 billion galaxies \citep{lsstsciencecollaboration2009lsstsciencebookversion}. With a detection rate of 2000 rings per million galaxies, we estimate there would be about 20 million true ring galaxies found in the entire dataset. With a current precision of $58.9\%$, applying this model would result in approximately 34 million detected ring galaxies, which includes 14 million false positives. This high number of false positives underscores the need for significant improvements in the model's accuracy to ensure reliable identification of ring galaxies and minimize false positives. Addressing this challenge is essential to maintain the viability of the model for extensive astronomical surveys.

With refined models, future sky surveys, such as the LSST, which will produce billions of images, can be systematically searched for ring galaxies. Applying this model to additional datasets will allow true positives to refine the model further and reduce the need for manual sorting. Existing sky surveys, which have yet to be explored, can be searched for thousands more ring galaxy candidates. Consequently, with less time spent obtaining an optimal sample of ring galaxies, their formation and evolution can be studied more in-depth with larger samples.
\subsection{Properties \& Analysis}

As the Galaxy Zoo 2 catalog of ring galaxies exhibited a lower star formation rate and extended into redder values compared to the rings identified by the CNN, they could be deduced to be further in their stages of galactic evolution. 

Additionally, as shown in Figure \ref{fig:mass}, the Galaxy Zoo 2 rings are less massive than the rings identified by the CNN. This could potentially indicate a bias of the model towards detecting younger, more massive galaxies compared to previous, human-compiled catalogs like Galaxy Zoo 2.

Looking at the r-band magnitudes of both samples of galaxies in Figure \ref{fig:magnitude}, the Galaxy Zoo 2 rings are seen to extend into higher magnitudes. However, as seen in Figure 
\ref{fig:magc}, which compares the r-band magnitudes for a random sample of rings from the Goddard-Shamir and the Galaxy Zoo 2 dataset, the magnitude distributions are around the same for both catalogs. Regardless of the fact that the model was trained on the Galaxy Zoo 2 sample of rings, the human-compiled Galaxy Zoo 2 catalog identified fainter galaxies than the machine learning approach could. This could mean that in future machine-classified catalogs of galaxies, fainter galaxies that humans could have identified may potentially be misclassified. A potential reason our model performed poorly on fainter galaxies could be the fact that the initial layers of the model were only trained on the galaxies simulated via GalFit, which were much brighter than the typical galaxy with no background stars or galaxies. The classification of fainter galaxies could be improved in the future by expanding our existing data augmentation methods, employing more advanced machine learning techniques, and including even fainter galaxies in the training dataset. 

The frequency histogram of redshifts for the newly identified rings is shown in Figure \ref{fig:rs}. The majority of the rings have redshifts between 0.02 and 0.12. To better understand the properties and evolution of ring galaxies, future research should focus on identifying higher redshift rings. By expanding the redshift range and obtaining a larger sample of high redshift rings, we can gain insights into the formation and evolution of these galaxies over a broader range of cosmic time.

The properties of the galaxies which were obtained in this analysis could possibly be used to conduct further investigations into these rings. The sample of 1967 newly classified rings, in addition, could be further investigated: spectroscopic data can be collected in future studies to further explore the dynamics of these galaxies. Additionally, this model could potentially be used to find rings in cosmological simulations like IllustrisTNG to better understand how they form and evolve as a whole.

\section{Conclusion} \label{sec:conclusion}

In this work, we investigate the usage of a convolutional neural network to detect large catalogs of ring galaxies from sky surveys. Rings are crucial to understand more about galaxy dynamics, and current catalogs of them are extremely limited.

Our network was first trained on a sample of 100,000 synthetic galaxies - simulated using {\tt\string GalFit} - and was then transfer learned to a sample of real galaxies consisting of 3117 rings, which was also augmented via a generative adversarial network (GAN). It was found that the Inception-ResNet V2 architecture achieved the best accuracy, and the best variation of transfer learning was when convolutional block B, C, and the output layers were retrained. The source code of the model, along with a catalog of the rings discovered, is made public at \url{https://github.com/harishk30/RingGalaxiesCNNAnalysis}.

After training, this network was applied to an unclassified sample of $\sim$950,000 galaxies, where it detected 1967 previously unclassified rings. However, the model's precision was limited to $58.9\%$, indicating the need for significant refinements before it can be reliably applied to larger sky surveys such as LSST.

After this, the properties of this sample were analyzed through available spectroscopic data and then compared to a sample of ring galaxies from Galaxy Zoo 2. The null hypothesis that the star formation rates of both the Galaxy Zoo 2 rings and the newly identified rings were from the same distribution could be rejected with a p-value of .002. Additionally, the probability of finding the Galaxy Zoo 2 rings at lower SSFRs was higher than the newly identified sample. The Galaxy Zoo 2 rings were also found to extend to bluer colors in the FUV - r color-mass diagram.

In the future, similar models could be used to extract large catalogs of rings from future sky surveys which will collect images of billions of more galaxies. However, before applying such models to extensive surveys, improving the model's precision is crucial to minimize false positives and ensure reliable classifications. Additionally, current sky surveys which have not yet been explored could be classified by this model to compile thousands of more existing rings which have yet to be classified. The sample of 1967 rings which were identified by this model could further be explored through conducting simulations and collecting more spectroscopic data, and insight about the formation and evolution of rings as a whole can be derived.

\section{Acknowledgments} \label{sec:akn}

We would like to thank Imad Pasha and Kate Allender for their help, without which this project could not have been completed. We also thank John Franklin Crenshaw for providing his comments on our paper. JBK acknowledges support from the DIRAC Institute in the Department of Astronomy at the University of Washington. The DIRAC Institute is supported through generous gifts from the Charles and Lisa Simonyi Fund for Arts and Sciences, and the Washington Research Foundation. This publication was supported by the Princeton University Library Open Access Fund.

\software{GalaxySim \citep{ghosh2020galaxy}, GALFIT \citep{2002}, Astropy \citep{astropy:2013, astropy:2018, astropy:2022}, Matplotlib \citep{Hunter:2007}, Numpy \citep{harris2020array}, Scipy \citep{2020SciPy-NMeth}, Pandas \citep{reback2020pandas}}

\appendix

\section{Artificial Neural Networks and Convolutional Neural Networks}
\label{sec:anncnn}

\begin{figure}[h]
\centering
\includegraphics[width=0.5\textwidth]{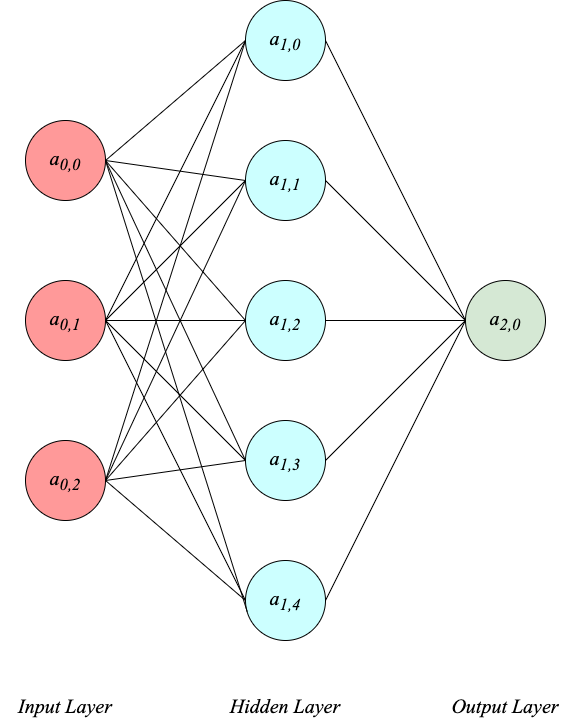}
\caption{An illustration of a simple MLP. This contains an input layer, only one hidden layer, and an output layer.}
\label{fig:mlp}
\end{figure}

A CNN is a type of artificial neural network (ANN) typically applied to analyze images. The simplest type of ANN is a multilayer perceptron (MLP), shown in Figure \ref{fig:mlp}. It consists of three layers - the input layer, the hidden layers, and the output layer. These layers consist of structures called neurons which are interconnected with those of the adjacent layers. These neurons each hold a value called their activation. The activations of the input neurons determine the activation of the subsequent layer, and this occurs until an activation for the output neurons are determined.

To determine the activations of the neurons in the subsequent layer, each connection between the previous layer and the next is assigned a weight and each neuron of the next is assigned a bias. Taking the first neuron in the second layer as an example, the weights of the connections are stored in a vector $w_{0} = \left(w_{0, 0}, w_{0, 1}, w_ {0, 2} \right)$, and the activations of the previous layer are stored in a vector $a_{0} = \left(a_{0,0}, a_{0, 1}, a_{0, 2} \right)$. The activation of this neuron is then given by

\begin{equation} \label{eq:C}
a_{1, 0} = \sigma\left(w_{0} \cdot a_{0} + b_{1,0}\right),
\end{equation}
where $b_{1,0}$ represents the bias of the neuron, and $\sigma$ represents the activation function, which is rectified linear unit (ReLU) in the CNN used to detect ring galaxies.

The weights and biases are first initialized with randomized values. However, this creates a network which performs extremely poorly on an input example. Thus, these values need to be modified through a process called backpropagation. For this, a cost function is needed, which describes how poorly a given network is performing. This is typically referred to as the loss, and in our CNN, binary cross-entropy loss is used, which is given by 

\begin{equation} \label{eq:D}
L = -\frac{1}{N}\sum_{i = 1}^{N} y_{i} \cdot \log \hat{y_{i}} + (1 - y_{i}) \cdot \log (1 - \hat{y_{i}}),
\end{equation}
where $N$ represents the number of outputs, $y_{i}$ represents the given class (1 or 0 in a binary case), and $\hat{y_{i}}$ represents the output probability. To compress the output neuron between 0 and 1, the sigmoid activation function has to be used.

To get the best possible network, $L$ needs to be minimized. In our CNN, this is done with stochastic gradient descent (SGD; \citealt{ruder2017overview}), which uses several iterations in mini-batches to find a relative minimum for the loss function through modifying the weights and biases. Simple SGD is given by

\begin{equation} \label{eq:E}
w_{t+1} = w_{t} - \alpha \cdot dW,
\end{equation}
where
\begin{equation} \label{eq:Z}
dW = \frac{\partial L}{\partial w_{t}}.
\end{equation}
In the equation, $t$ is the time step, $w$ is the weight of a certain connection and $\alpha$ is the learning rate, which controls how quickly the weights are changed. The same concept is also applied to modify bias. However, SGD can further be improved upon with the Adam optimizer \citep{kingma2017adam}, which uses an adaptive learning rate, meaning that different parameters have different learning rates. This allows for faster convergence on a relative minimum of the loss function. Adam combines two other optimizers, Momentum and RMSProp. Adam is given by 

\begin{equation} \label{eq:G}
w_{t+1} = w_{t} - \alpha\frac{m_{t}}{\sqrt{s_{t} + \epsilon}},
\end{equation} 
where 

\begin{equation} \label{eq:H}
m_{t} = \beta_{1} \cdot m_{t-1} + (1 - \beta_{1}) \cdot dW
\end{equation} and

\begin{equation} \label{eq:O}
s_{t} = \beta_{2} \cdot s_{t-1} + (1 - \beta_{2}) \cdot dW^{2}.
\end{equation} 
$\beta_{1}$, $\beta_{2}$ and $\epsilon$ are constants which typically have values of 0.9, 0.999 and $10^{-8}$ respectively. The same concept can also be applied to find the correct value for bias. 

A CNN (\citealt{Fukushima1980}; \citealt{lecun2015deep}) is a type of ANN which is typically used to classify images. A CNN differs slightly from the basic MLP in that there are stages of pre-processing to the input where key features are extracted. First, a convolutional layer takes a 3D block of data as an input. For images, this is a 3D matrix of their pixel values. A kernel, a matrix smaller than the input, is initialized with certain values. These values can be learnt through several iterations with SGD. This kernel moves across the initial image, and the dot product between the kernel and the smaller patch of the initial image is stored in a new matrix referred to as a feature map. An activation function, ReLU in this case, can be applied along with the convolution to modify the feature map. A convolutional layer is then followed by a max-pooling layer, which slides a kernel across the feature map and stores the largest value of each patch into a newly created feature map. These layers are alternated until a sufficient abstraction of the initial image is achieved. The outputted feature maps are then inputted into a series of perceptron layers, similar to the MLP, and are reduced to an output containing a prediction as to what class the input image belongs to. The parameters for these layers are also learnt through SGD as seen with the ANN before. 

\begin{figure} 
\plotone{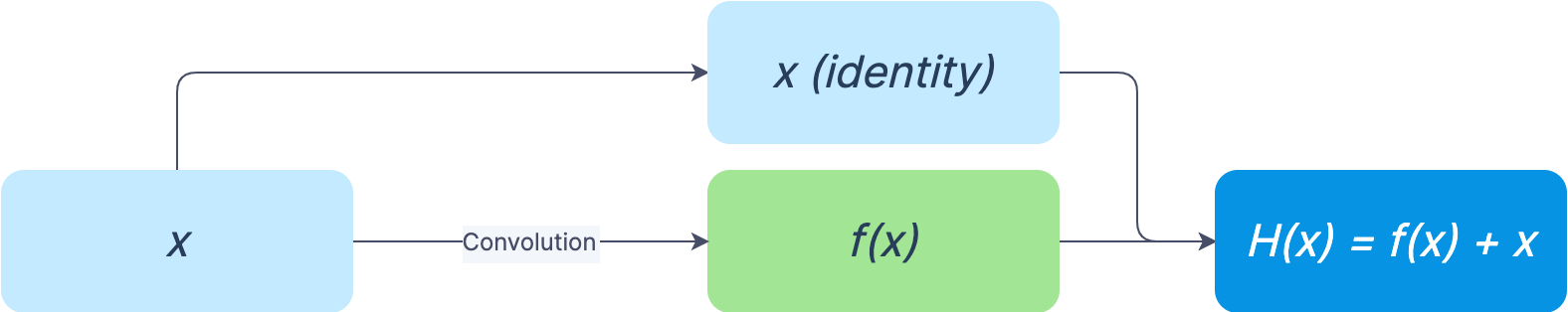}
\caption{An illustration of the residual connections of ResNet. The identity of the initial function, $x$, is added to the feature map after convolution, $f(x)$, to produce the output.}
\label{fig:res}
\end{figure}

Typically, more layers can be added to a network to increase its accuracy. For extremely deep networks, the feature maps cannot be further downsized. Then, identity mapping is required to add more layers, where convolutions are performed with a kernel which will output a feature map with the same dimensions. However, for a standard CNN, this is difficult to learn, and results in the degradation problem, where the accuracy resulting with the addition of more layers initially increases, but falls after reaching a maximum. 

To combat this problem, deep residual networks (ResNets; \citealt{he2015deep}) were created. As illustrated in Figure \ref{fig:res}, these networks take the initial feature map, and store its identity in a variable, $x$. Then, a convolution is applied to the initial feature map to produce $f(x)$. To produce the final feature map, $H(x)$, $f(x)$ is combined with $x$. These networks also help alleviate the vanishing gradient problem, where $dW$ becomes too small to make significant changes to the weights. ResNets are created by stacking many of these types of layers.

To make deeper networks, inception modules \citep{szegedy2014going} can be used as well. These modules use a concatenation of 3x3 max pooling and 1x1, 3x3 and 5x5 convolutions within a single layer to create a feature map. Inception modules offer the advantage of allowing for the creation of a larger network and the extraction of features at various scales, thus possibly learning more features.

However, with an extremely deep network comes the risk of being too attached to the training data, known as overfitting. Thus, several methods were implemented to ensure that the risk of overfitting was very minimal on the network. The first of these methods was regularization, which adds an extra penalty term to the loss function to ensure that it does not assume extreme values. In this network, $L_{2}$ regularization \citep{cortes2012l2} was used. The modified loss function with $L_{2}$ regularization is given by
\begin{equation} \label{eq:I}
L_{new} = L + \frac{1}{N} \frac{\lambda}{2}\sum w_{i}^2,
\end{equation}
where the sum of the squared differences between the real and target weights is multiplied by the reciprocal of the number of outputs, $N$, and $\frac{\lambda}{2}$, where $\lambda$ is the $L_{2}$ regularization constant, which is set at $0.0001$ in the network. Additionally, data augmentation, where the initial data is transformed via several methods and expanded, was used. A scale of the images by a factor of 0.1, a shear by a factor of 1.2, a zoom by a factor of 0.25 and rotations in the range $0^\circ$ to $360^\circ$ were used.


\bibliography{RingBib}{}
\bibliographystyle{aasjournal}

\end{document}